\documentclass[12pt,letterpaper]{article}
\pdfoutput=1

\usepackage{amsmath,amssymb,calc, amsthm}
\usepackage{graphicx}
\usepackage[nosort]{cite}
\usepackage{epsfig,psfrag}

\usepackage{ulem}
\usepackage{color}
\definecolor{burgundy}{rgb}{0.8,.2,.2}

\makeatletter
\renewcommand\section{\@startsection {section}{1}{\z@}%
                                   {-3.5ex \@plus -1ex \@minus -.2ex}
                                   {2.3ex \@plus.2ex}%
                                   {\normalfont\large\bfseries}}
\renewcommand\subsection{\@startsection{subsection}{2}{\z@}%
                                     {-3.25ex\@plus -1ex \@minus -.2ex}%
                                     {1.5ex \@plus .2ex}%
                                     {\normalfont\bfseries}}
\makeatother

\theoremstyle{plain}

\theoremstyle{definition}

\let\non\nonumber

\let\a=\alpha
\let\b=\beta

\let\s=\sigma

\let\S=\Sigma
\let\Th=\Theta

\newcommand{\del}{\partial}
\newcommand{\delbar}{\bar{\partial}}

\def\one{^{(1)}}

\newcommand{\bea}{\begin{eqnarray}}
\newcommand{\eea}{\end{eqnarray}}
\newcommand{\be}{\begin{equation}}
\newcommand{\ee}{\end{equation}}
\newcommand{\bma}{\begin{pmatrix}}
\newcommand{\ema}{\end{pmatrix}}

\newcommand{\hlf}{\frac{1}{2}}

\newcommand{\Z}{{\mathbb Z}}
\newcommand{\R}{{\mathbb R}}

\newcommand{\F}{{\mathbb F}}
\newcommand{\PP}{{\mathbb P}}
\newcommand{\CC}{{\mathbb C}}

\newcommand{\cC}{{\cal C}}
\newcommand{\cA}{{\cal A}}

\newcommand{\B}{{\cal B}}

\newcommand{\Om}{\Omega}

\renewcommand{\O}{\operatorname{O}}

\newcommand{\La}{\Lambda}
\newcommand{\la}{\lambda}

\newcommand{\G}{\Gamma}
\newcommand{\e}{\epsilon}

\newcommand{\dd}{\delta}

\newcommand{\f}{\psi}

\newcommand{\D}[1]{\ensuremath{\mathrm{D}#1}}

\newcommand{\C}[1]{$(\ref{#1})$}

\def\IZ{\relax\ifmmode\mathchoice
{\hbox{\cmss Z\kern-.4em Z}}{\hbox{\cmss Z\kern-.4em Z}}
{\lower.9pt\hbox{\cmsss Z\kern-.4em Z}} {\lower1.2pt\hbox{\cmsss
Z\kern-.4em Z}}\else{\cmss Z\kern-.4em Z}\fi}
\def\IR{\relax{\rm I\kern-.18em R}}

\def\one{{\hbox{ 1\kern-.8mm l}}}

\newlength{\bredde}
\def\slash#1{\settowidth{\bredde}{$#1$}\ifmmode\,\raisebox{.15ex}{/}
\hspace*{-\bredde} #1\else$\,\raisebox{.15ex}{/}\hspace*{-\bredde}
#1$\fi}

\newsavebox{\zzzbar}
\sbox{\zzzbar}
  {\setlength{\unitlength}{0.9em}
  \begin{picture}(0.6,0.7)
  \thinlines
  \put(0,0){\line(1,0){0.6}}
  \put(0,0.75){\line(1,0){0.575}}
  \multiput(0,0)(0.0125,0.025){30}{\rule{0.3pt}{0.3pt}}
  \multiput(0.2,0)(0.0125,0.025){30}{\rule{0.3pt}{0.3pt}}
  \put(0,0.75){\line(0,-1){0.15}}
  \put(0.015,0.75){\line(0,-1){0.1}}
  \put(0.03,0.75){\line(0,-1){0.075}}
  \put(0.045,0.75){\line(0,-1){0.05}}
  \put(0.05,0.75){\line(0,-1){0.025}}
  \put(0.6,0){\line(0,1){0.15}}
  \put(0.585,0){\line(0,1){0.1}}
  \put(0.57,0){\line(0,1){0.075}}
  \put(0.555,0){\line(0,1){0.05}}
  \put(0.55,0){\line(0,1){0.025}}
  \end{picture}}

\def\Im{{\rm Im\, }}

\newcommand{\alt}[1]{\left\{\begin{array}{ll} #1 \end{array}\right.}

\newcommand{\ena}{\end{eqnarray}}
\newcommand{\beqa}{\begin{eqnarray}}
\newcommand{\eeqa}{\end{eqnarray}}

\def\G{\Gamma}

\def\cD{{\cal D}}

\renewcommand{\b}{\beta}
\newcommand{\g}{\gamma}

\def\d{{\rm d}}

\newcommand{\ibar}{{\bar \imath}}
\newcommand{\jbar}{{\bar \jmath}}
\newcommand{\thbar}{{\bar \theta}}
\newcommand{\Dbar}{{\bar D}}

\newcommand{\labar}{{\bar \lambda}}

\newfont{\goth}{ygoth.tfm scaled 1200}                   

\def\a{\alpha}
\def\b{\beta}
\def\e{\epsilon}
\def\th{\theta}
\def\f{\phi}
\def\g{\gamma}

\def\j{\psi}

\def\o{\omega}

\def\s{\sigma}

\def\D{\Delta}
\def\F{\Phi}
\def\G{\Gamma}

\def\L{\mathcal{L}}
\def\O{\Omega}

\def\S{\Sigma}

\renewcommand{\O}{{\mathcal{O}}}

 \numberwithin{equation}{section}

\def\1{{(1)}}
\def\2{{(2)}}
\def\3{{(3)}}

\def\1{{\bf 1}}
\def\a{{\alpha}}

\def\B{{\mathcal B}}

\def\CC{{\mathbb C}}

\begin{document}
\begin{titlepage}

\begin{center}

{June 13, 2012}
\hfill         \phantom{xxx}  EFI-12-09

\vskip 2 cm {\Large \bf  Novel Branches of $(0,2)$ Theories}
\vskip 1.25 cm {\bf  Callum Quigley$^{a}$\footnote{cquigley@uchicago.edu}, Savdeep Sethi$^{a}$\footnote{sethi@uchicago.edu} and Mark Stern$^{b}$\footnote{stern@math.duke.edu}}\non\\
\vskip 0.2 cm
 {\it $^{a}$Enrico Fermi Institute, University of Chicago, Chicago, IL 60637, USA}\non\\ \vskip 0.2cm
$^{b}${\it Department of Mathematics, Duke University, Durham, NC 27708, USA\non\\}

\end{center}
\vskip 2 cm

\begin{abstract}
\baselineskip=18pt

We show that  recently proposed linear sigma models with torsion can be obtained from unconventional branches of conventional gauge theories. This observation puts models with log interactions on firm footing.  If non-anomalous multiplets are integrated out, the resulting low-energy theory involves log interactions of neutral fields. For these cases, we find a sigma model geometry which  is both non-toric and includes brane sources. These are heterotic sigma models with branes. Surprisingly, there are massive models with compact complex non-K\"ahler target spaces, which include brane/anti-brane sources. The simplest conformal models describe wrapped heterotic NS5-branes. We present examples of both types.

\end{abstract}

\end{titlepage}


\section{Introduction}
\label{intro}

Perturbative string theory involves the study of two-dimensional quantum field theory. A large class of field theories are obtained from two-dimensional non-linear sigma models for which there is a natural geometry given by the target space manifold. The aim of this project is to consider new branches of linear sigma models with $(0,2)$ supersymmetry and explore their associated geometries.

Our motivation is to understand a class of linear sigma models with torsion proposed in~\cite{Quigley:2011pv, Blaszczyk:2011ib}. For earlier interesting work on torsional linear theories, see~\cite{Adams:2006kb, Adams:2009av}. The novelty of these models is the inclusion of field-dependent Fayet-Iliopoulos (FI) couplings. The field-dependence is via log interactions of scalar fields in a superpotential coupling of schematic form:
\be\label{sketch}
S = \int d^2x \, d\th^+ \, \log (\S) \Upsilon.
\ee
Unlike four-dimensional ${\cal N}=1$ theories, a superpotential coupling in a $(0,2)$ theory is fermionic. The field $\S$ is a conventional chiral superfield, while the Fermi superfield $\Upsilon$ contains the complexified gauge-field strength: $\theta^+ \left( F_{01}+iD\right)$. Again unlike in four dimensions, supersymmetry pairs the topological theta-angle coupling appearing in~\C{sketch}\ with the $D$-term potential rather than with the gauge coupling. This is the primary reason an interaction like~\C{sketch}\ leads to a change in the sigma model geometry and to the appearance of $H$-flux.

It is very reasonable to be concerned by the appearance of a non-polynomial interaction like the logarithm of~\C{sketch}\ in the definition of a theory. Whenever one sees log interactions, it is  natural to suspect that these interactions can be generated from a theory without non-polynomial interactions by integrating out massive fields. For at least a sizable class of models, we will show that this is indeed the case. This puts the existence of the quantum theory with the logarithm on much firmer footing since these models naturally appear from a conventional framework.

 If the $\S$ field appearing in~\C{sketch}\ is charged under an abelian gauge symmetry then gauge invariance is broken at the classical level. This violation of gauge invariance can be compensated by
a one-loop quantum anomaly, giving an intrinsically quantum consistent theory described in~\cite{Quigley:2011pv}. If the argument of the logarithm is gauge-invariant then the models are classically consistent.  With multiple abelian gauge fields, it is even possible to have charged logarithms and still preserve classical gauge invariance using $A$-$V$ couplings. Gauge invariant models of this type first appeared in an interesting attempt to prove $(2,2)$ mirror symmetry~\cite{Morrison:1995yh}.

Surprisingly, there are many interesting classically gauge-invariant intrinsically $(0,2)$ models, even without $A$-$V$ couplings. We will explore a class of such models in this work.
In terms of field theory dynamics, $(0,2)$ theories are highly reminiscent of four-dimensional ${\cal N}=1$ gauge theories while $(2,2)$ theories are more akin to four-dimensional ${\cal N}=2$ theories.

In $(0,2)$ theories, there are generically many branches in the moduli space aside from a standard Higgs or Coulomb branch. The other branches involve directions in what is traditionally viewed as the gauge bundle moduli space, but which can actually appear on the same footing as the directions conventionally making up the sigma model geometry. These unconventional branches, to be described below, are the home of compactifications with both fluxes and branes.

\subsection{The basic idea} \label{basicidea}

To see how these branches arise, start with a conventional $(2,2)$ model viewed from a $(0,2)$ perspective. Consider a $(2,2)$ $U(1)$ vector multiplet. Viewed from a $(0,2)$ perspective, this multiplet contains two gauge superfields $(A, V_-)$ and a chiral superfield $\Sigma$, whose superspace expansions are given in Appendix~\ref{superfieldconventions}, along with our superspace conventions. This vector multiplet can be obtained by dimensionally reducing an ${\cal N}=1$ four-dimensional vector multiplet. The neutral chiral superfield $\Sigma$ captures the two scalars that arise in this reduction.

On the other hand, a $(2,2)$ chiral multiplet with charge $Q$ decomposes into a pair $(\Phi, \G)$ of $(0,2)$ superfields consisting of a $(0,2)$ chiral multiplet $\F$, and an almost chiral Fermi multiplet $\G$ satisfying~\cite{Witten:1993yc}
\be
\bar{\mathfrak{D}}_+ \G = \sqrt{2} E.
\ee
The lowest component of the superfield $\F$ is a complex scalar $\f$, while the lowest component of $\S$ is a complex scalar $\s$.

For a Fermi superfield that comes from a $(2,2)$ multiplet, there is a prescribed relation with $E = \sqrt{2} Q \Sigma \Phi$. For general $(0,2)$ models, $E$ can be a more interesting function of all the chiral superfields and we will exploit this freedom. In fact, the $E$ degree of freedom is as rich as a conventional superpotential in terms of physics, but far less well-explored. To see this, note that the total bosonic potential for a $(0,2)$ gauge theory takes the form
\be\label{bosonic}
V_{\rm bos} = {1\over 2 e^2} |D|^2 +|J|^2 +  |E|^2.
\ee
There are two sets of holomorphic data entering~\C{bosonic}, which are on equal footing. The first is the conventional superpotential, $J$, and the second is the choice of $E$.

Now consider an illustrative case. Take $G=U(1)$ gauge theory coupled to $n+1$ $(2,2)$ chiral multiplets of charge $Q_i$. We will distinguish the first $n$ fields $\F^i$ from the $(n+1)$ chiral field, which we denote $P$. In the absence of any superpotential, the classical target space for this theory is a toric variety; for the case $Q_i=Q_P=1$, the space is $\PP^n$.
From the perspective of a $(0,2)$ model, there are two bosonic potentials. The first is the $D$-term potential,
\be\label{dterm}
\sum_i  Q_i |\phi^i|^2 + Q_P |P|^2= r,
\ee
where $r$ is the FI parameter. The second is the potential associated to the left-movers $\G^i$ given by $|E|^2$.

Our immediate interest is in geometry rather than the left-moving gauge-bundle, so we will not worry about the question of whether the left-movers define a (semi-)stable or unstable bundle in the Higgs phase. We do, however, require vanishing of the one-loop gauge anomaly for consistency of the theory; this is a quadratic condition on the charges. Let us choose
\be
E^P = \S P, \qquad E^i= 0.
\ee
This is a $(0,2)$ preserving deformation of the $(2,2)$ model for which $E^i=  \sqrt{2} Q_i \S \F^i$. The classical vacuum structure is found by solving~\C{dterm}\ and the condition $|E^P|^2=0$.

Unlike the $(2,2)$ model where solving $|E|^2=0$ implies $\s=0$, there is now a classical branch where $(p=0, \s\neq 0)$. In the $(2,2)$ model, there is a Coulomb branch where $(\phi^i =p=0, \s \neq 0)$, but it is not a classical zero energy branch except for the choice $r=0$. There are, however, vacuum solutions on this branch when one-loop effects are included~\cite{Witten:1993yc}. The central role of the $\s$ field and this Coulomb branch was originally realized in the large $n$ analysis of the $(2,2)$ $\PP^n$ model~\cite{DAdda:1982eh}.  Our new $(0,2)$  branch emanates from the usual Higgs branch at a locus of complex co-dimension one.\footnote{Ilarion Melnikov has amusingly termed these branches ``horns'' sticking out of the conventional Higgs branch.} This is a kind of Higgs branch in which a ``bundle'' direction, $\s$, has become part of the geometry! Such branches are generic in $(0,2)$ models.

For  large $\s$, we can include the leading quantum effect by integrating out the massive field $P$ giving a modified $D$-term,
\be
\sum_{i}  Q_i |\phi^i|^2 - N \log |\s| = {\hat r}.
\ee
The coefficient $N={1 \over 2\pi} Q_P$, while ${\hat r}$ is a renormalized FI parameter. This is a model of the kind described in~\cite{Quigley:2011pv}, but with a log interaction involving a neutral field. We see that at least a class of those models arise as novel branches of more conventional $(0,2)$ theories.

\subsection{Future directions and an outline}

Our goal is to study geometries of this type. As we will see, there are both compact and non-compact examples. It is natural to suspect that these models might be conformal for suitable charges, and we will investigate that possibility. Clearly, there are many generalizations. The examples considered here only involve neutral log interactions. Even in this setting, it is interesting to understand how $A$-$V$  couplings, described in~\cite{Quigley:2011pv}, are generated by integrating out fields in models with multiple $U(1)$ factors.

All of these cases provide natural generalizations of toric geometry, and understanding these geometries is going to be interesting. The example described in section~\ref{basicidea}\ takes  the form of a weighted projective space with its K\"ahler class fibered over the $\s$-plane. The point $\s=0$ is special since there is a new massless degrees of freedom; namely, the integrated out $p$-field. This point will correspond to the brane source.

Perhaps the most exciting future direction is the case of quantum geometries where the $\S$ field of~\C{sketch}\ is charged. The classical geometries in these cases are not only non-toric but not even complex! For example, non-complex spheres like $S^4$ naturally emerge in the simplest models. This strongly indicates the need to take into account new physics from the light anomalous chiral fermions. Otherwise, it would appear that supersymmetry is broken.

It is very natural to ask whether these quantum theories can also be found as branches of conventional $(0,2)$ gauge theories. This indeed appears to be true and comes about as follows: in the example of section~\ref{basicidea}, the scalar $p$ -field becomes massive on the new branch along with a gauge non-anomalous combination of left and right-moving fermions. All these fields are integrated out leaving a gauge-invariant model. However, this model is still too closely wedded to its $(2,2)$ origins. There is no reason to consider $E$-couplings which include just a neutral chiral superfield $\S$. One could just as well consider the following $E$-coupling for a Fermi field $\G$,
\be
E = \S_1 \S_2,
\ee
where both $\S_1$ and $\S_2$ are charged. Suppose $\S_1\neq 0$ so $\S_2$ masses up. Necessarily, the associated combination of massed up left and right-moving fermions is now gauge anomalous. We expect a pion-like coupling involving $\log(\S_1)$ which reproduces the anomaly of these massive fermions, together with additional quantum corrections. This should be the right framework to determine the low-energy description of the quantum compactifications described in~\cite{Quigley:2011pv}. This direction, which requires more subtle computations, is currently being explored~\cite{toappearcallum}.

Our paper is organized as follows: in section~\ref{prelim}, we describe the basic setup and the conditions that must be satisfied by the metric, flux and dilaton if they are to describe a heterotic string vacuum.  In section~\ref{solutions}, we construct the perturbative infra-red (IR) geometry and fluxes for the unconventional branches of the $(0,2)$ gauged linear sigma model (GLSM). We describe the obstructions to conformality by studying the leading quantum corrections in the GLSM. The resulting condition for a conformal model is very much in accord with our expectations from the IR geometry.

Section~\ref{exam}\ contains a collection of examples including massive compact geometries constructed with branes and anti-branes. These are complex non-K\"ahler spaces. For example, $S^5 \times S^1$ emerges from our construction as a nice smooth case. We also present non-compact conformal models which describe wrapped NS5-branes. Lastly, we discuss some obstructions to building compact conformal models via complete intersections.

\section{Preliminaries}\label{prelim}

Let us begin by assembling some facts about the structure of the world-sheet solutions and the constraints on heterotic space-time solutions. Our superspace and superfield conventions can be found in Appendix~\ref{superfieldconventions}.

\subsection{The gauge group action}

Consider a  $G=U(1)^r$ abelian gauge theory.   Coupled to these gauge fields are $n$ chiral superfields $\Phi^i$ with charges $Q^a_i$. The bosonic lowest components of $\Phi^i$ are denoted $\phi^i$. Under a gauge transformation with parameters $\Lambda^a$,
\be \label{gaugeaction}
\Phi^i \rightarrow e^{i \Lambda^a Q^a_i} \Phi^i.
\ee
The gauge fields are arranged into gauge superfields $A^a$ and $V^a_-$ with $a=1, \ldots r$. The corresponding field strength is a fermonic superfield $\Upsilon^a$.
We will also include a set of $m$ neutral chiral fields $\S^\a$. We will restrict our attention to a single $\S$ field in constructing examples, but let us keep the number general for this preliminary discussion. The novelty in the construction of the theory is to consider superpotential interactions,
\be\label{FIparam}
S_{\rm log} =  {i\over 4} \int d^2x d\th^+ \, N_\a^a \log (\S^\a) \Upsilon^a + c.c.,
\ee
which modify both the $D$-term constraints and introduce $H$-flux into the resulting geometries.\footnote{For convenience, we are changing the original sign convention of~\cite{Quigley:2011pv}. With this new sign convention, brane-like solutions correspond to positive $N$ while anti-brane-like solutions correspond to negative $N$. } There can also be gauge-invariant non-logarithmic couplings in~\C{FIparam}, but we will focus on the log case.

We will assume $2\pi N_\a^a\in { \Z }$. This quantization condition is certainly consistent with models where the logs are obtained by integrating out charged fields as described in section~\ref{basicidea}; it might be possible to relax this condition for models which are not obtained from this UV completion. There is a $D$-term constraint for each gauge factor,
\be \label{dterms}
\sum_i Q^a_i |\phi^i|^2 - N_\a^a \log |\s^\a| = r^a.
\ee
The solution of the $D$-term constraints is a surface $W_{r,Q,N} \subset \CC^{n+m}$. The geometric moduli space is the further quotient by the global gauge group
\be X_{r,Q,N}  = W_{r,Q,N}/G.\ee
The basic defining data are the charges $Q_i^a$, the integers $N_\a^a$, and the FI parameters $r^a$.
We will assume integral $Q_i^a$.

Unlike the models discussed~\cite{Quigley:2011pv}, here the $\S$ fields are neutral. This means that the action is gauge invariant and there is no need to introduce $A$-$V$ couplings or quantum anomalies to restore gauge invariance. This allows us to explore the essential features of theories with log interactions without many of the complications that arise when the $\S$ fields are charged.

If all $N_\a^a=0$, this combinatorial data describes a toric variety constructed via symplectic reduction in the following way: consider the algebraic torus $\left(\CC^\ast \right)^n$ acting on  $\Phi$ by
\be
\Phi^i \rightarrow  \la^i \Phi^i, \qquad (\la^1, \ldots \la^n) \in\left(\CC^\ast \right)^n.
\ee
The quotient by $G$ removes the compact part of a $\left(\CC^\ast \right)^r$ action specified by the charges $Q^a_i$ and the action~\C{gaugeaction}. If all $N_\a^a=0$ then we can find a unique solution to the $D$-term constraints~\C{dterms}\ in the orbit of the $\left(\CC^\ast \right)^r$ action acting on any sufficiently generic choice of $\Phi^i$. This fixes the scaling symmetry in $\left(\CC^\ast \right)^r$. We can therefore view solving the $D$-term constraints (which determine $W$) and quotienting by $G$ (which determines $X$) as gauge-fixing $\left(\CC^\ast \right)^r$. The moduli space is  a toric variety characterized by a fan.

If all $N_\a^a$ are not zero then the $D$-term constraint can have multiple solutions. The resulting space cannot be viewed as gauge-fixing a $\left(\CC^\ast \right)^r$ action. Rather, the existence of multiple solutions changes the topology of the space. For example, non-toric spheres can appear. Inclusion of the log interactions provides a very natural generalization of toric geometry.

\subsection{The $(0,2)$ metric and flux}

We expect classically gauge invariant models derived from $(0,2)$ superspace to have complex target manifolds. The metric $G$ for the target manifold determines a $(1,1)$ fundamental form $J$ via
\be
J_{i\bar{j}} = i G_{i \bar{j}}.
\ee
In turn, the fundamental form determines the torsion via
\be \label{susyreln}
H = i (\bar\partial - \partial) J.
\ee
The Hodge decomposition of $H$ contains no $(3,0)$ or $(0,3)$ components as a consequence of $(0,2)$ supersymmetry. These constraints are automatically satisfied for models constructed in $(0,2)$ superspace. Let us ignore the left-moving Fermi degrees of freedom, and focus on the geometry and flux. The superspace Lagrangian for a non-linear sigma model takes the form
\be \label{action}
\L = -{i\over4}\int\d^2\th^+\ \left(K_i(\F,\bar{\F})\del_-\F^i - K_\ibar(\F,\bar{\F})\del_-\F^\ibar\right).
\ee
The defining data is a $(1,0)$ form $K = K_i\d\f^i$ with complex conjugate  $K^*=K_\ibar\d\f^\ibar$. The $1$-form $K$ is the analogue of the K\"ahler potential found in $(2,2)$ theories. The target space fields are determined by $K$,
\be \label{GandB}
G_{i\jbar} = K_{(i,\jbar)} \qquad {\rm and} \qquad B_{i\jbar} = K_{[i,\jbar]}.
\ee
The  $(0,2)$ analogue of a K\"ahler transformation is
\be \label{K1}
K(\F,\bar{\F}) \rightarrow K(\F,\bar{\F}) + K'(\F)
\ee
where $ K'(\F)$ is any holomorphic $(1,0)$-form. These transformations leave the physical couplings of~\C{GandB}\ invariant. Furthermore, a shift in $K$ of the form
\be \label{K2}
K \rightarrow K +i\,\del U,
\ee
for any real-valued function $U$, amounts to a $B$-field transformation $\dd B = i\del\delbar U$. This shifts the Lagrangian~\C{action}\ by a total derivative, and therefore is also a symmetry of the action.

\subsection{Conditions for a space-time supersymmetric solution}

Most $(0,2)$ sigma models are not conformal and will not provide solutions to the heterotic space-time equations of motion. The heterotic conditions for a space-time supersymmetric solution were derived from supergravity in~\cite{strominger-torsion}, and considered from a pure spinor perspective in~\cite{Andriot:2009fp}. We would like to understand the local conditions on $K$ required for a space-time solution with Minkowski space-time. In a perturbative $\alpha'$ expansion, there are no four-dimensional solutions of de Sitter or anti-de Sitter type, with or without space-time supersymmetry~\cite{Green:2011cn, Gautason:2012tb}.

For K\"ahler metrics, the condition for a supersymmetric Minkowski solution is Ricci-flatness and requires solving a Monge-Amp\`ere equation,
\be\label{monge-ampere}
\del\delbar\log\det\left( G\right) = 0,
\ee
for the target space metric  $G$ expressed in holomorphic coordinates.
For backgrounds with NS-flux, the conditions are more involved because of the $H$-field and associated varying dilaton. For $(2,2)$ models with flux and varying dilaton, a generalized Monge-Amp\`ere equation constraining the generalized $(2,2)$ K\"ahler potential (which includes semi-chiral fields) was described in~\cite{Hull:2010sn}.

In this analysis, we are considering $(0,2)$ models which are classically gauge-invariant. Cancellation of the one-loop gauge anomaly between the left and right-moving sectors implies that the Bianchi identity for $H$ is trivial at leading order in $\alpha'$,
\be
dH=0  + O(\alpha').
\ee
The non-closed components of $H$ are $O(\alpha')$; if the curvature scale of $G$ is small, we should therefore find consistent solutions to the space-time equations of motion at the level of heterotic supergravity. We note that our metrics and fluxes should satisfy:
\bea
 R_{MN}+2 \nabla_M \nabla_N
\varphi -{1\over 4} {H}_{MAB} {{H}_N}^{AB} &=& O(\alpha'), \label{eom}\\
 d \left( e^{-2
\varphi} \star {H}\right) &=& O(\alpha'^2), \\
\nabla^2\varphi -2\nabla_M\varphi\nabla^M\varphi + \hlf|H|^2 &=& O(\a'),
\eea
where $\varphi$ is the string dilaton.\footnote{We have assumed that we are in the critical dimension for the heterotic string. Otherwise, the dilaton equation of motion would have an additional term proportional to the central charge of the theory.}  In addition to the equations of motion, we expect space-time supersymmetry to be unbroken by the metric, flux and dilaton. It might be broken by the choice of gauge bundle but that is an effect higher order in $\alpha'$. Ignoring the gaugino constraint, space-time supersymmetry requires the existence of a Killing spinor $\e$ satisfying
\bea
 \label{gravitinovar}\delta \Psi_M  &=& \left(\nabla_M -{1\over 4} H_M \right)\epsilon=0, \\
  \label{dilvar} \delta \lambda &=& \left(  /\!\!\!
\partial \varphi  -{1\over 2}/\!\!\! \!{ H} \right) \epsilon=0.
\eea
The first condition~\C{gravitinovar}\ requires $SU(n)$ structure for a complex $n$-dimensional target manifold. This implies the existence of a nowhere vanishing holomorphic top form $\Omega$ satisfying
\be
d\left( e^{-2\varphi} \Omega \right)=0.
\ee
Note that condition~\C{susyreln}\ is automatically satisfied for any model with $(0,2)$ supersymmetry. Space-time supersymmetry also implies a constraint on $J$:
\be
d\left( e^{-2\varphi}  J^{n-1} \right) =0.
\ee

\subsubsection{An alternative characterization}\label{alternate}

The constraints on the geometry, flux and dilaton of a $(0,2)$ solution can be elegantly encoded in properties of the torsionful connection
\be \Om_M^{(-)}=\Om_{M}-\hlf H_M, \ee
with $\Om_M$ the usual spin connection; see, for example, Appendix A of~\cite{Gillard:2003jh}\ or~\cite{Ivanov:2000ai}. Note that the torsionful affine connection contains a relative sign:
\be
\G_{MN}^{(\pm)P} = e^P_A\left(\del_M e^A_N + e^B_N\Om_M^{(\pm)A}{}_B \right) = \G^P_{MN} \mp\hlf H^P{}_{MN}.
\ee
The two Killing spinor equations~\C{gravitinovar}\ and~\C{dilvar}\ imply the existence of an integrable complex structure that is covariantly constant with respect to $\Om^{(-)}$.  A Hermitian manifold satisfying this property is called K\"ahler with torsion (KT). Covariant constancy of the complex structure implies the constraint~\C{susyreln}, which can be re-written as follows,
\be\label{rewrite}
\G_{i\jbar}^{(-)k} = \G_{\ibar\jbar}^{(-)k}= 0,
\ee
so that $\Om^{(-)}$ has $U(n)$ holonomy.
The gravitino equation~\C{gravitinovar}\ implies that the holonomy of $\Om^{(-)}$ is actually in $SU(n)$ rather than $U(n)$. This holds iff  ${\cal R}^{(-)} = d\o^{(-)}=0$, where
\be
\o_i^{(-)} = i\G_{ij}^{(-)j} - i\G_{i\jbar}^{(-)\jbar} = 2i G^{j\bar{k}}\del_j G_{i\bar{k}} - i G^{j\bar{k}}\del_i G_{j\bar{k}}
\ee
is the connection on the canonical bundle induced by $\Om^{(-)}$. This is a natural torsional generalization of a Calabi-Yau space. Condition~\C{rewrite}, which is a rewriting of~\C{susyreln}, follows automatically from $(0,2)$ superspace whether the model is conformal or not.  Imposing conformal invariance requires $SU(n)$ structure.

To solve the dilaton supersymmetry constraint~\C{dilvar}, it is useful to introduce the Lee form of a KT manifold defined by,
\be \xi = -2i\delbar^\dagger J, \label{definexi}\ee
where $\delbar^\dagger$ is the adjoint of $\delbar$.\footnote{There is a factor of $2$ in~\C{definexi}\ because the Lee form appears in the modification of the K\"ahler identities for a non-K\"ahler space. See page 307 of~\cite{Demailly_2009}.}
The components of $\xi$ are determined in terms of $G$,
\be
\xi_i = iH_{ij\bar{k}}J^{j\bar{k}} = G^{j\bar{k}}\left(\del_i G_{j\bar{k}} - \del_j G_{i\bar{k}}\right).
\ee
 In terms of the Lee form, the dilatino equation~\C{dilvar}\ becomes,
\be \label{xitriv}\xi =2\partial\varphi,\ee
with $\varphi$ real. KT manifolds with exact Lee forms are conformally balanced. An explicit check that conformally balanced KT manifolds with $SU(n)$ structure solve the supergravity equations~\C{eom}\ can be found~\cite{Ivanov:2000ai}.

One might ask under what conditions $SU(n)$ structure implies a solution of the dilaton constraint. To relate the two constraints, note that \be\label{omegaminus}
\o^{(-)} = i(\del-\delbar)\log\det G -2i\xi + 2i\bar{\xi}.
\ee
The condition of $SU(n)$ structure then requires,
\be\label{flatness}
 \del\xi - \delbar\bar{\xi} + \delbar\xi -\del\bar{\xi} +\del\delbar\log\det G=0.
\ee
This condition is the generalization of the Monge-Amp\`ere equation~\C{monge-ampere}\ to KT manifolds. Following~\cite{strominger-torsion}, we can examine the $(0,2)$ part of this equation which implies
\be
\delbar \bar\xi = 0.
\ee
At least on a space with $h^{(0,1)}=0$, we can conclude that $\bar\xi = 2\delbar \varphi$ for some complex $\varphi$. It remains to show that $\varphi$ can be chosen real. It is not unreasonable to expect this to be true in fairly general circumstances for compact manifolds.\footnote{It might be possible to show this for compact spaces with $h^{(0,1)}=0$ by modifying the argument of~\cite{strominger-torsion}, where a simply-connected space is assumed. There are two complications that need to be addressed. First: on a non-K\"ahler space, $\sum_{p+q=n} h^{p,q} \geq b_{n}$ (see, for example~\cite{Becker:2003yv}) so simply-connected is not sufficient to guarantee exactness of the Lee form; however, assuming $h^{(0,1)}=0$ is good enough for $\delbar$ triviality. The second complication is that the $\square_{\partial}$ and $\square_{\bar\partial}$ Laplacians differ by linear differential operators that depend on $\xi$ (see~\cite{Demailly_2009}). This complicates the original proof of~\cite{strominger-torsion}\ that ${\rm Im}(\varphi)$ is constant on a compact space. } Our examples will be both non-compact and non-simply-connected so we will need to examine what can be said about the Lee form in each case.

When~\C{xitriv}\ is satisfied with a real $\varphi$, we can rewrite the generalized Monge-Amp\`ere equation~\C{flatness}\ as follows:
\be
\del\delbar\log\left(e^{-4\varphi}\det G\right) = 0.
\ee
In summary, a KT manifold with $SU(n)$ structure and a (de Rham) exact Lee form provides a supersymmetric heterotic string solution.

\section{Non-Linear Geometries from Linear Models}
\label{solutions}

We are going to construct metrics and fluxes for a non-linear sigma model starting from a $(0,2)$ GLSM. The procedure we will follow is to ignore the gauge kinetic terms (which formally vanish in the infra-red limit) and integrate out the abelian gauge-fields. The result is a non-linear sigma model determined by a metric and flux.
Consider a model with field-dependent  FI-terms,
\be\label{f}
{i\over4}\int\d^2x\d\th^+\, N^a\log(\S)\Upsilon^a + c.c.,
\ee
as motivated in the introduction and section~\ref{prelim}.
It is natural to define the field-dependent variables,
\bea
R^a(\s) = r^a + N^a\log|\s|, \qquad \Th^a(\s) = N^a\Im\log\s - {\th^a\over2\pi} ,
\eea
which include possible constant FI parameters $(r^a, \th^a)$. We use $T^a$ to denote the complexified total FI parameter:
\be
T^a = t^a + iN^a\log\s = iR^a - \Th^a; \qquad t^a = ir^a + {\th^a \over 2\pi}.
\ee

The most effective way to determine the induced metric and flux is to first find the induced $K$ in superspace.  The $V_-^a$ superfields only appear as Lagrange multipliers that enforce the superfield constraints,
\be \label{constraint}
\sum_i Q^a_i|\F^i|^2 e^{2Q^b_iA^b}  = R^a(\S).
\ee
The constraint~\C{constraint}\ determines the superfields $A^a$ implicitly in terms of $(\F, \bar{\F})$ and $R(\S)$. We will use the notation $A^a$ for both the lowest scalar component of the superfield as well as the superfield itself. Hopefully, the usage is clear from context. Equation~\C{constraint}\ is a generic polynomial in $e^{2A_a}$ so we can only find explicit solutions for $A_a$ for simple charge assignments. However, we can get surprisingly far just knowing that~\C{constraint} is satisfied.

The first thing to notice is that the $A^a$ are not globally defined functions, but are sections of some set of line bundles $\L^a$ over the target space. To see this, consider a simple case with $G=U(1)$. In a patch $U_{(\a)}$ where $\f^\a\neq0$, we can define gauge-invariant coordinates
\be
Z^i_{(\a)} = \left(\f^i\right)\left(\f^\a\right)^{-Q_i/Q_\a}. \label{coords}
\ee
On the intersection $U_{(\a)}\cap U_{(\b)}$, the coordinates then transform as follows:
\be
Z^i_{(\a)} = Z^i_{(\b)} \left(Z^\a_{(\b)}\right)^{-Q_i/Q_\a}.
\ee
However, since the right hand side of~\C{constraint}\ is invariant, it follows that
\be
A_{(\a)} = A_{(\b)} + {1\over Q_\a} \log\left|Z^\a_{(\b)}\right|,
\ee
so $A$ is not globally defined. However, note that $\del A$ behaves like a connection on $\L^a$, while the curvature two-form $\del\bar{\del}A$ \textit{is} globally defined.  All of these quantities will play a role in the following discussion.

We can express $K$ in terms of the $A^a=A^a(|\F|^2,R)$ superfields,
\bea\label{KfromGLSM}
K_i = \bar{\F}^i e^{2Q^a_iA^a} - 2i \Th^a \del_i A^a, \qquad K_\s = \bar{\s} - 2i\Th^a\del_\s A^a.
\eea
Differentiating~\C{constraint}\ yields the useful relations
\be
\del_i A^a = -\bar{\f}_i\D^{ab}Q_i^a e^{2Q_i^c A^c},\qquad \del_\s A^a = \D^{ab}\del_\s R^b,
\ee
where we introduce the quantity
\be\label{delta}
\D^{ab} = {\del A^b\over \del R_a} = \left(2\sum_i Q^a_i Q^b_i |\f^i|^2 e^{2Q_i^c A^c}\right)^{-1}.
\ee
We follow the convention that $\bar{\f}_i \equiv \dd_{i\jbar}\bar{\f}^\jbar$, $\f_\jbar \equiv \dd_{i\jbar} \f^i$. These relations allow us to determine the induced sigma model metric
\bea
G_{i\jbar} = e^{2Q_i^cA^c}\left(\dd_{i\jbar}-2\bar{\f}_i\f_\jbar Q^a_i \D^{ab} Q_j^b e^{2Q_j^cA^c}\right), \label{metric1} \qquad G_{\s\bar{\s}} = 1+ {N^a\D^{ab}N^b \over2|\s|^2},
\eea
and $B$-field
\be\label{bfield1}
B = -2i\Th^a\del\bar{\del}A^a.
\ee
One should not worry too much about the detailed form of these solutions because they will be modified under RG flow; however, we do expect the RG flow to preserve the coarse, topological features.

As one of these features, note that the metrics take the form of a warped product over the $\s$-plane, with no off-diagonal mixing between the fiber and base. In addition, the $B$-field roughly takes the form $\Th^a F^a$ where $F^a\sim i\del\bar{\del}A^a$ is the curvature of the line bundle $\L^a$. Even though the $F^a$ are closed, the field-dependence of $\Th^a$ means that $B$ is \textit{not} closed, and there is a non-zero flux $H\sim d\Th^a\wedge F^a$. In particular, the components of $H$ are
\bea\label{H1}
H_{i\s\bar{\s}} = -i \left( \del_\s \Th^a \del_{i\bar\s} A^a - \del_{\bar\s} \Th^a \del_{i\s} A^a \right), \qquad
H_{i\s\jbar} = -i \del_\s \Th^a \del_{i\bar j} A^a,
\eea
which requires use of the relation $\del_{i\jbar}A^a = -\hlf\del_{R_a} G_{i\jbar}$. It is natural to identify
\be J^a_{i\jbar}(R) =i \del_{R_a} G_{i\jbar}, \ee
with the generators of $H^2$ for the toric fiber.


\subsection{Obstructions to conformality}
\label{obs}
The geometric data one obtains directly from a GLSM construction almost never gives $SU(n)$ structure on the nose. It is reasonable to assume that as long as the cohomology class $\left[{\cal R}^{(-)}\right]$ is trivial, the metric will flow to the one with $SU(n)$ structure in the IR. Similar reasoning is used in the standard Calabi-Yau case.
As we saw in section~\ref{alternate}, once we have a metric with ${\cal R}^{(-)}=0$ the associated fundamental form $J$ determines the $H$-flux as well as the Lee form $\xi$ which, if exact, fixes the dilaton.

We therefore expect a $(0,2)$ sigma model to define heterotic string background if the class $\left[{\cal R}^{(-)}\right]$ is trivial.
 The GLSM provides a choice of coordinates on the target space. From the induced couplings given in~\C{metric1}\ and~\C{bfield1}, we can determine the components of the induced Lee form in these distinguished coordinates,
\be\label{lee}
\xi_i = \del_i\log G_{\s\bar{\s}},\qquad \xi_\s = \del_\s\log\det G_{i\jbar}.
\ee
Now $G_{\s\bar{\s}}$ is a globally defined object, but $G_{i\jbar}$ is not. This observation combined with the form of $\o^{(-)}$ given in~\C{omegaminus}\ implies a single obstruction; namely, that
\be \del \bar\del \log\det G_{i\jbar} \ee
be a trivial class. Equivalently, the original Monge-Amp\`ere equation~\C{monge-ampere}\ for the fiber metric should be satisfied at the level of cohomology. However, this is just the familiar requirement that the toric fibers have vanishing first Chern class, or in terms of GLSM data that $\sum_i Q_i^a=0$. We provide a proof of this fact in Appendix~\ref{proof}\ in order to demonstrate that this familiar result continues to hold even for this more general class of non-K\"ahler solutions. We should stress that this condition on the charges is for the residual theory obtained after integrating out $(P, \G_P)$ multiplets as described in the introduction~\ref{basicidea}. We will construct some non-compact conformal examples satisfying this constraint in section~\ref{noncompactconformal}.  

We should be a little careful about the claims of the previous paragraphs. Although it sounds very reasonable, it has not yet been proven that the triviality of $\left[{\cal R}^{(-)}\right]$ implies the existence of an $SU(n)$ structure metric. An analogue of Yau's proof of the Calabi conjecture for KT manifolds is needed to show that vanishing of the cohomological obstruction is sufficient. This kind of result is a little less interesting for $(0,2)$ models compared with $(2,2)$ models because we expect ``most'' compact KT metrics to involve small volumes of order the string scale, like the solutions of~\cite{Dasgupta:1999ss, Becker:2009df}.

 Actually, the metric and flux for a conformal model with a large volume limit will not satisfy just the supergravity equations of motion, but the equations of motion including $\alpha'$ corrections. What is really needed for these theories is a statement about RG flow that generalizes the analysis of~\cite{Nemeschansky:1986yx}\ to $(0,2)$ models. This would involve a classification of the cohomological obstructions that could appear under renormalization. Again for most compact models, an analysis that goes beyond $\alpha'$ perturbation theory is desirable.

The last issue is whether the dilaton equation can be solved. We must ensure that the Lee form is exact with a real potential. This is non-trivial to see starting with GLSM data. The GLSM expression for the induced Lee form given in~\C{lee}\ is not even closed. However, the form is completely determined by the metric. Under RG flow, we expect the metric to flow to one appropriate for a conformal field theory and the IR Lee form should be determined by that metric. The GLSM Lee form is not exact but it is given by gradients of real functions. In the simplest conformal model, we will give evidence that the Lee form actually becomes exact with a real potential by studying the large $Q_P$ limit.

We also note that the GLSM expression~\C{lee}\ has no components proportional to $d\,{\rm Im}\left( \log\s \right)$ which generates $H^1$ for our examples. It seems plausible that RG flow will not produce a component non-trivial in cohomology, but a sharp argument is desirable.
It is worth contrasting this situation with the well-studied $S^3\times S^1$ SCFT, where solutions with $SU(2)$ structure exist, but the Lee form is not exact with a component along the $S^1$ direction. These theories do not define good string backgrounds unless $S^1$ is replaced with ${\mathbb R}$ trivializing the Lee form. The result is the NS5-brane background. We present the details of this example in Appendix~\ref{S3xS1}.



\subsection{Quantum corrections} \label{quantum}

In the introduction, we explained how a theory with log couplings can arise from a standard GLSM. Let us now study how this  happens in greater detail. The idea is to use $E$-couplings to generate a mass for a chiral and Fermi superfield pair ($P,\G_P)$, along a branch where $\langle\s\rangle\neq0$. Since we wish to assign canonical dimension 0 to $\S$, we must introduce a mass scale for the $E$-coupling. In a $(2,2)$ theory, $\S$ is part of the vector multiplet so this scale would naturally be set by the two-dimensional gauge coupling $e$. However, in a $(0,2)$ theory we are free to introduce another mass scale, which we call $m_0$. Then the $E$-couplings we want to consider are
\be
E_P = m_0 \S P,\qquad E_i=0.
\ee
When $\s\neq0$, the scalar field $p$ (along with its right-moving fermionic superpartner $\psi_P$ and the left-moving fermion $\gamma_P$) becomes massive with a mass $m=m_0|\s|$. Below the scale $m$, we should integrate out the superfields $P$ and $\G_P$ which generates the field-dependent FI couplings~\C{f}, where $N^a={Q_P^a\over 2\pi}$. In particular, the bare FI parameters get modified as follows,
\be\label{renormr}
r_0^a \rightarrow r^a + N^a \log|\s|,
\ee
where $r^a = r_0^a + N^a \log(m_0/\La)$ is the renormalized FI parameter and $\La$ is a UV cutoff scale~\cite{Witten:1993yc, Morrison:1994fr}.

In addition to these FI couplings, integrating out $(P,\G_P)$ also modifies the kinetic terms of the vector multiplet:
\be\label{kinetic}
\L_{D,F} = \hlf\left({\dd^{ab}\over e_a^2} + 2\pi {N^a N^b \over m^2} +\ldots\right)\left(D^a D^b + F_{01}^a F_{01}^b \right),
\ee
where the ellipses denote terms that are more suppressed than $O(1/m^2)$. This is a standard computation that can be found, for example, in~\cite{Hori:2003ic}. In the IR limit where we send $e_a^2,m_0^2\rightarrow\infty$, we see that the these kinetic terms decouple  provided $\s$ is not too small. In actuality, since we are discussing a quantum mechanical theory in two dimensions, there is no well-defined expectation value for $\s$. Rather there is a branch with $|\s| \neq 0$ which can be studied in a Born-Oppenheimer approximation as long as $|\s|$ is sufficiently large.

In the approximation where we neglect the kinetic terms~\C{kinetic}, $D^a$ and $A^a_\mu$ act as Lagrange multipliers. The constraint of sufficiently large $\s$ should not be too surprising since it just means we are in a regime where we can trust integrating out $(P,\G_P)$ at one-loop. Near $\s=0$, there can be large quantum corrections but we will still be able to study the basic features of our solutions near this point.

There is one more quantum correction induced by integrating out $(P,\G_P)$, which is a correction to the $\S$ kinetic terms. In the large $m_0$ limit,
\be\label{sigmakinetic}
\L_\s = \left(1+{1\over 8\pi |\s|^2} + \ldots\right)|d\s|^2,
\ee
with additional corrections suppressed by ${1/ m_0^2}$. Again, this is only reliable away from $\s=0$. This correction has an important effect since the induced sigma-model metric is significantly modified:
\be\label{renormG}
G_{\s\bar{\s}} = 1 + {1\over2|\s|^2}\left({1\over4\pi} + N^a\D^{ab} N^b\right).
\ee

Finally, since we are considering the theory below the scale $m$, we should also integrate out the high-energy modes of the rest of the fields. As in the usual case, the main effect of this integration is to modify the FI parameters in a way determined by the sum of the charges. Our previous expression for the FI parameters~\C{renormr}\ becomes
\be
r^a + N^a \log|\s| +{1\over2\pi}\left(\sum_i Q_i^a\right)\log\left(\mu\over\La\right),
\ee
where $\mu$ is some IR cutoff scale that we need to introduce since the fields $\f^i$ are massless. In this Wilsonian effective action, no further $\s$-dependent corrections are possible because the FI couplings are controlled by holomorphy. In particular, when $\sum_i Q^a_i=0$ the FI parameters do not run below the scale $m$ and the theory can flow to a non-trivial conformal point. It is reassuring to see the same condition we found in section~\ref{obs}\ for the low-energy sigma-model also emerges here from RG flow of the GLSM.

To summarize, the picture we find goes as follows: far above the scale $m_0$, we have a standard GLSM with chiral fields $\left(\F^i,P,\S \right)$ with charges $(Q_i^a,Q_P^a,0)$. The FI parameters run according to $Q_P^a +\sum_i Q_i^a$. As we run down to the scale $m_0$, we integrate out $P$ which generates the couplings~\C{f}\ as well as the corrections to the $\s$ kinetic terms given in~\C{sigmakinetic}. Below the scale $m_0$, we have a GLSM with log interactions and the running of the FI parameters is controlled by $\sum_i Q_i^a$. When the sum of these charges vanishes, $r^a$ does not run and the theory can flow to a conformal fixed point.

In the deep IR, the theory flows to a conformal sigma-model whose target is a toric space fibered over the $\s$-plane. The sizes of various two-cycles in the fibers are controlled by the field-dependent quantities $R^a=r^a+N^a\log|\s|$. In general, one combination of the $r^a$ parameters can always be absorbed into the zero-mode of $\s$, along with its corresponding $\th^a$. In this sense, the log interactions remove moduli from the sigma model.

When all $R^a$ are large, the non-linear sigma model geometry should provide a reliable guide to the physics. However, in general there will be regions where some or all of the $R^a$ become small, or even negative. In these regions, one expects another description (like an orbifold SCFT) to be the appropriate description. The correct description can, nevertheless, be determined from the GLSM starting point. This is very much like the phase structure of~\cite{Witten:1993yc}, but with some of the FI parameters promoted to dynamical fields. In some regions, the description is geometric while in others non-geometric. The entire structure glues together to form a single quantum field theory.

\section{Examples}\label{exam}

\subsection{A non-compact massive model}
\label{example1}
The case with multiple $U(1)$ factors can become complicated quickly. Our strategy will be to search for examples of interesting spaces with the number of $U(1)$ factors small. The first interesting case involves just a single $U(1)$ factor. This is a case which should allow us to isolate the essential physics that differentiates these models from conventional branches of $(0,2)$ theories.

Let us begin in the UV with a collection of $n+1$ charge $+1$ chiral fields $\F^i$, along with our distinguished field $P$ with a charge $Q_P$ that can be positive or negative. In the positive case, the conventional $(0,2)$ Higgs branch is a weighted projective space. In the negative case, the Higgs branch is the total space of a line bundle over projective space.

\begin{figure}[ht]
\begin{center}
\[
\mbox{\begin{picture}(230,220)(80,40)
\includegraphics[scale=1]{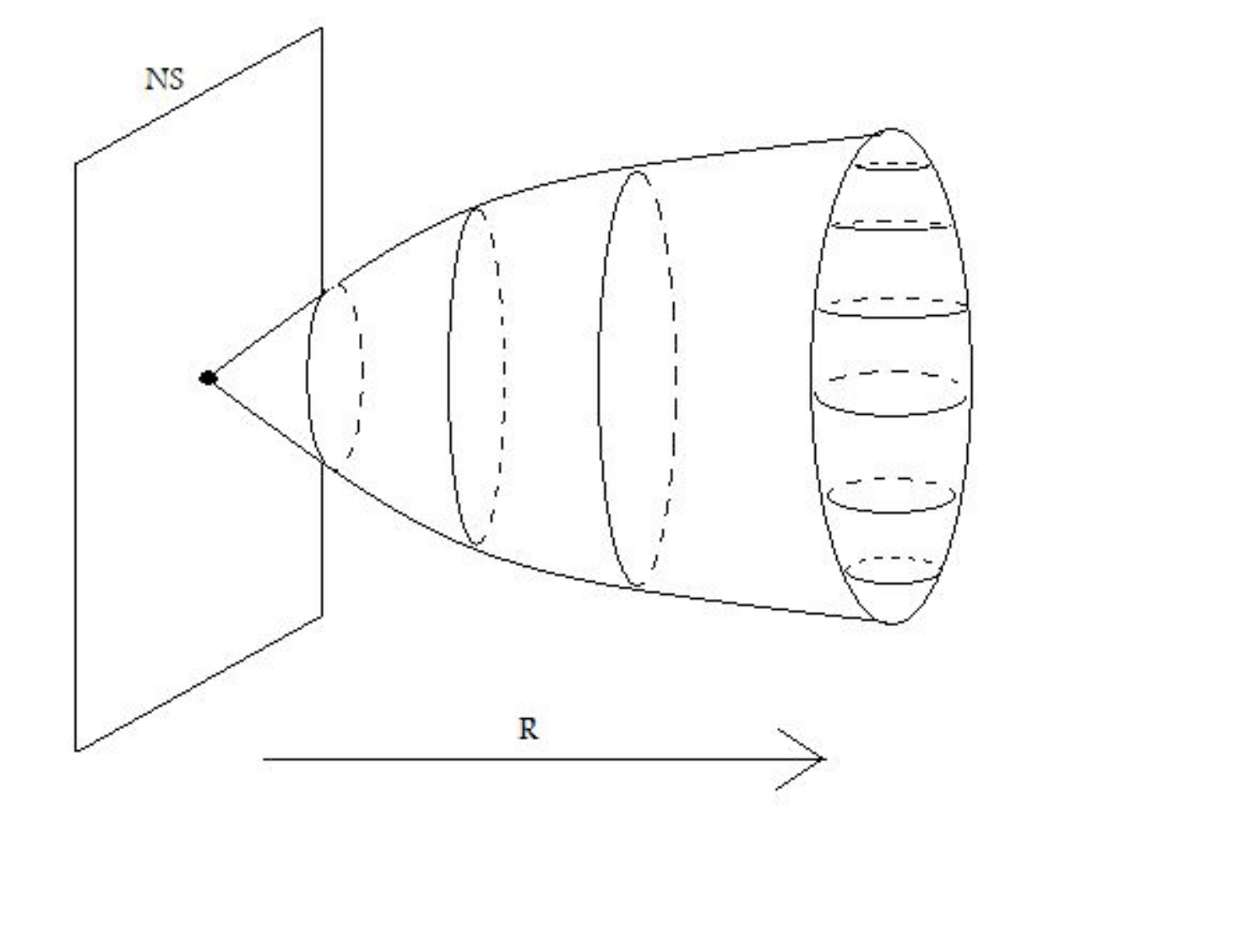}
\end{picture}}
\]
\vskip 0.2 in \caption{\it A sketch of the brane geometry in the massive $(0,2)$ model. } \label{figure1}
\end{center}
\end{figure}

Now imagine moving to the branch in which the $\Sigma$ field becomes massless while $P$ masses up, as described in section~\ref{quantum}.  The $D$-term constraint on this branch is given by,
\be\label{Dterm1}
\sum_{i=1}^{n+1}  |\phi^i|^2 = R(\s) = r +  N \log |\s|
\ee
where $N={1 \over 2\pi} Q_P$. The space looks like projective space $\PP^n$ (parameterized by the $\f^i$) fibered over the $\s$-plane. The allowed range of $\s$ is fixed by the positivity of $R$, and it depends on the sign of $N$,
\bea
|\s| \in \alt{ (e^{-r/N},\infty), & N>0 \\  (0,\,e^{r/|N|}), & N<0 }
\eea
Interestingly, the effective description is always valid for $N>0$ since $|\s|$ is bounded away from 0, but the $N<0$ models can access that region (corresponding to $R\rightarrow\infty$) where there are large quantum corrections. We will therefore focus on the case $N>0$. It will turn out to be natural to work with the complex variable
\be T=iR -\Th=t +iN\log\s \ee
rather than $\s$.  Note that $T$ has periodicity $T\simeq T+2\pi N$.

Since $\sum_i Q_i=n+1$, we already know that the theory has a mass gap.\footnote{Actually, we should be more careful about whether there is really a mass gap if the $E$-couplings are set to zero. It is possible that the left-moving fermions with $E=0$ flow to a chiral current algebra in the IR. This happens, for example, for the Schwinger model with flavors. We wish to thank Ilarion Melnikov for explaining this possibility. Here we focus on whether the right-moving sector, which characterizes the geometry, flows to a SCFT.  Whether this is possible depends on the sum of the $U(1)$ charges.} Nevertheless, many of the key features that show up in all models with log interactions appear in this  example. This class is particularly nice since $A$, solving~\C{constraint}, has a very simple form
\be
A= \hlf \log\left(R\over|\f^i|^2\right),
\ee
which allows us to determine $\D= \del_R A = {1\over2 R}$. Plugging these expressions into~\C{metric1}-\C{H1}, and including the quantum correction~\C{renormG}, we find the induced metric and $B$-field
\bea
ds^2 &=& R g_{FS}(\f) + \left({e^{2(R-r)/N} + 1/8\pi\over N^2} +{1\over4R}\right)|dT|^2, \label{metricPn}\\
B &=& \Th \left(J_{FS}(\f) + i{dT\wedge d\bar{T}\over4 R^2}\right),
\eea
where $g_{FS}$ and $J_{FS}$ are the Fubini-Study metric and K\"ahler form on $\PP^n$. Notice that the radius of the $\Th$ circle diverges at the boundaries $R=0$ and $R\rightarrow\infty$, but never vanishes in the interior. In particular, the size of $S^1_\Th$ does not vanish.

The fundamental two-form and flux of the total space are,
\bea
J &=& R J_{FS}(\f) +i \left({e^{2(R-r)/N} + 1/8\pi\over N^2} +{1\over4R}\right)dT\wedge d\bar{T}, \\
H &=& d\Th\wedge J_{FS}(\f),
\eea
and, as a consistency check, it is easy to see that these satisfy the SUSY relation:
\be
H = dB = i(\bar{\del}-\del)J.
\ee

Notice that near $R=0$, the $\PP^n$ fiber is shrinking to zero size. In particular the two-cycle class $\cC$, dual the K\"ahler form $J_{FS}$, is pinching off. This trivializes $\cC$ in the total space. Even though $\cC$ is trivial in homology, when we integrate $H$ over $\cC\times S^1_\Th$ we get a non-zero result:
\be
\int_{\cC\times S^1_\Th} H = 2\pi N = Q_P,
\ee
at any value of $R$. This indicates that there is a collection of $Q_P$ NS-brane sources located at $R=0$. In fact, we can use this structure as the definition of NS-branes in massive $(0,2)$ theories.  We have depicted these geometries in figure~\ref{figure1}.

Note that we expect new physics to become important at $R=0$ since, according to the $D$-term constraint~\C{Dterm1}, all the $\f^i=0$. This point is therefore a Coulomb branch since all charged fields vanish, but the physics at this point can still be gapped if there is a non-zero theta-angle, much like the conventional Coulomb branches of $(2,2)$ theories~\cite{Coleman:1976uz, Witten:1993yc}.
To get a better understanding of what is happening near this Coulomb point, we write the metric near $R=0$ as
\be
ds^2 = R g_{FS}(\f) +{| dT|^2\over4R}. \label{metricR0}
\ee
We could also obtain this metric from~\C{metricPn}\  by taking $N\rightarrow\infty$. Notice that the FI parameter $r$ no longer appears in the metric, so the metric and $B$-field have no tunable moduli in this limit. This ``near-horizon" metric has a few interesting equivalent forms. First by writing $R=e^U$, we find
\be
ds^2 = e^U\left(g_{FS}(\f) + {1\over4}ds^2_{PD} \right),
\ee
where $ds_{PD}^2 = dU^2 + e^{-2U}d\Th^2$ is the metric on the Poincar\'e punctured disk. So this space is conformal to a product of two symmetric spaces.

Another equivalent form is as a cone over $\PP^n\times S^1$, though in a peculiar way. Letting $R=\tilde{R}^2$ gives
\be
ds^2 = d\tilde{R}^2 + \tilde{R}^2 g_{FS}(\f) + {d\Th^2\over4\tilde{R}^2}.
\ee
The radius of the $S^1$ goes to zero as $\tilde{R}\rightarrow\infty$ but blows up at the origin, while the size of $\PP^n$ varies in the opposite way. Furthermore, the shrinking $\PP^n$ leads to a conical singularity at the origin.\footnote{The exception is the case $n=1$ where we find a collapsing $\PP^1\cong S^2$ and $\tilde{R}=0$ is a smooth point in $\R^3$. }

To further explore the nature of this singularity, we can perform a T-duality along the $\Th$ direction. It helps to first extract the factor of $N$ from $\Th$ so that it has canonical periodicity $2\pi$. The $T$-dual space then turns out to be the orbifold $\CC^{n+1}/\Z_{Q_P}$, with metric
\be
\widetilde{ds}^2 =d\tilde{R}^2 + \tilde{R}^2 g_{FS}(\f) + {4\tilde{R}^2\over N^2}\left|d\tilde{\Th}^2-N A(\f)\right|^2,
\ee
where $dA=J_{FS}$ and $\tilde{\Th}$ is the coordinate of the dual circle. This orbifold can be viewed as a cone over an $S^1$ bundle over $\PP^n$, where the twist charge of this fibration precisely matches the NS-brane charge $Q_P=2\pi N$ in the original space. The dual space does not contain any $H$-flux.

This picture is in agreement with the basic duality relating NS5-branes with $A$-type $ALE$-spaces. Indeed the precise field theory to which these models flow will depend strongly on the choice of left-moving sector. In particular, whether instantons are localized at the orbifold point.

\begin{figure}[ht]
\begin{center}
\[
\mbox{\begin{picture}(280,280)(60,30)
\includegraphics[scale=0.45]{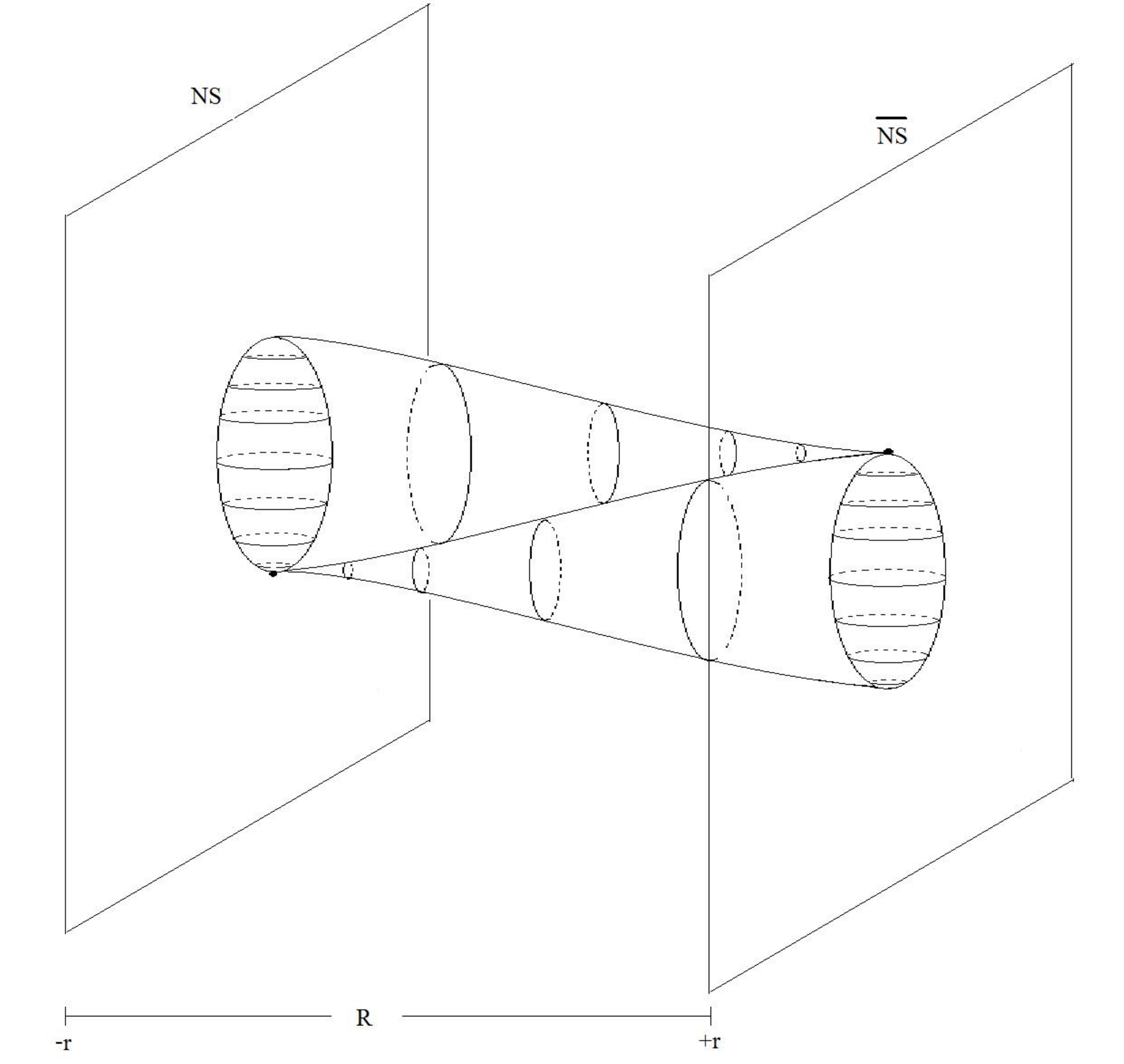}
\end{picture}}
\]
\vskip 0.2 in \caption{\it A sketch of $S^5 \times S^1$ constructed by gluing together branes and anti-branes. } \label{figure2}
\end{center}
\end{figure}

\subsection{A compact massive model}\label{compact}

It should be clear from the previous example that a model with log interactions and a single $U(1)$ gauge group will always be non-compact, since nothing prevents $R\rightarrow\infty$. An easy way to get compact models of this type is to include a second gauge group, and choose $D$-terms so that $\s$ is bounded. As a nice class of examples, consider two sets of chiral fields: $n+1$ chirals $\F^i$ with charges $(1,0)$ and $m+1$ chirals $\tilde{\F}^k$ with charges $(0,1)$. As before, we integrate out a field $P$ to generate the log interactions, but now with charges $(Q_P,-Q_P)$ under the two gauge groups. For definiteness, let us assume $Q_P$, and hence $N$, is positive. This leads to the following set of $D$-term equations:
\bea\label{twodterms}
\sum_i |\f^i|^2 &=& r^1 + N\log|\s|, \\
\sum_k |\tilde{\f}^k|^2 &=& r^2 -N\log|\s|.
\eea
After quotienting by $U(1)^2$, the solution space takes the form $\PP^n\times \PP^m$ fibered over the $\s$-plane, except now the range of $\s$ is bounded:
\be
e^{-r^1/N}\leq |\s| \leq e^{r^2/N}.
\ee
Notice that $|\s|$ is always bounded away from $0$. In fact, $|\s|$ can be made large by tuning $(r^1, r^2)$ which makes the inclusion of just the leading one-loop quantum corrections quite reliable. One of the two projective spaces collapses at each of the $|\s|$ boundaries resulting in a conical singularity and a Coulomb branch for the relevant $U(1)$. Only for the case $n=m=1$ is the total space smooth. For this special case, the $\PP^1\times\PP^1$ fibration over the $\s$-plane is actually $S^5\times S^1$ with the $S^1$ factor corresponding to the $\Th$ circle. Note that this is a complex space!

Another useful way to think about these spaces is to take another combination of gauge groups. Consider the $U(1)$ diagonal in the original $U(1)\times U(1)$ group and its complement. For these combinations, the $D$-term equations become
\bea
&&\sum_i |\f^i|^2 + \sum_k |\tilde{\f}^k|^2 = 2r \equiv  r^1 +r^2,  \\
&&\sum_i |\f^i|^2 - \sum_k |\tilde{\f}^k|^2 = 2R \equiv r^1- r^2 +2N\log|\s|,
\eea
where the factors of 2 have been chosen for later convenience. Now we see that from the original two FI parameters, only the sum $2r=r^1+r^2$ has any physical meaning; it fixes the size of a $\PP^{n+m+1}$ inside the total space. The other combination of FI parameters just gets absorbed into the variable $R$, which now takes values in the range $[-r,r]$. Similarly for the $\theta$-angles where $2\th=\th^1+\th^2$ measures the $B$-field threading the $\PP^{n+m+1}$, while $2\Th = 2N\Im\log\s +\th^2-\th^1$ parametrizes the free circle.

The metric and flux for this class of models is a straightforward generalization of the previous cases:
\bea
ds^2 &=& \left(r+R\right)g_{FS}(\f) + \left(r-R\right)g_{FS}(\tilde{\f}) + \Bigg({e^{(2R-r^1+r^2)/2N}  +{1/8\pi} \over N^2} +\cr && {r/2\over r^2-R^2}\Bigg) |dT|^2,  \\
H &=& d\Th\wedge \left(J_{FS}(\f) - J_{FS}(\tilde{\f})\right).
\eea
Denoting the non-trivial two-cycle classes of $\PP^n$ and $\PP^m$ by $\cC$ and $\tilde{\cC}$, we can again integrate $H$ to get finite results
\be
\int_{\cC\times S^1} H = +Q_P,\qquad \int_{\tilde{\cC}\times S^1} H = - Q_P,
\ee
even though the classes $\cC$ and $\tilde{\cC}$ are trivial in the total space. This indicates that these spaces contain a stack of $Q_P$ branes at $R=-r$ and $Q_P$ anti-brane sources at $R=+ r$. These spaces are basically two copies of the non-compact example of section~\ref{example1}\ corresponding to brane and anti-brane sources glued together to form a compact geometry. We have sketched this geometry in figure~\ref{figure2}.

\subsection{Non-compact conformal models}\label{noncompactconformal}\label{noncompactconformal}

We now turn to the construction of a class of conformal solutions. In this subsection, we only consider cases with $G=U(1)$. As in the preceding non-conformal examples, when we focus on a rank one gauge group nothing bounds the value of $R$, so these models will all be non-compact. While the arguments of section~\ref{obs}\ give us confidence that we only need to impose $\sum_i Q_i=0$ to guarantee that the IR non-linear sigma model is conformal, we have little hope of following the RG flow to determine the exact geometric data that characterizes the IR solution. We do expect coarse features, like the NS-brane charge and the topological structure of the metric, to remain invariant. We can study this data in specific models.

The analysis also simplifies considerably if we examine a large charge limit. For a single $U(1)$ gauge group, the components of the induced Lee form are
\be \label{lee1}
\xi_\s = \del_\s \log \D,\qquad \xi_i = \del_i \log\left(1+ {1/4\pi + N^2 \D\over2|\s|^2}\right).
\ee
This form is not exact, nor do we expect it to be exact in the UV; however, it has no component along the ${\rm arg}(\s)$ direction so there is no immediate obstruction preventing flow to something exact in the IR. If we consider the $N\rightarrow\infty$ limit, we \textit{do} find an exact Lee form, namely $\xi=d\log\D$. In this limit, we can try to identify the dilaton by setting $e^{2\varphi}=\D$. The precise form of $\D$ depends on the charge assignments, and $\D$ itself may be subject to renormalization.  However, this does provide an indication that the GLSM solutions simplify in the large $N$ limit, and begin to exhibit features expected in the IR conformal field theories.

The simplest cases to consider are just extensions of those in section \ref{example1}, with $n+1$ chiral fields of charge $+1$, together with one chiral field $\Phi^0$ with charge $-n-1$. Note that the addition of a negatively charged field means that $R$ is no longer positive definite. In the UV, for $R>0$ the resulting spaces will be fibrations of the local Calabi-Yau geometry $\O(-n-1)\rightarrow\PP^{n}$ over the $\s$-plane, with $H$-flux supported on a $2$-cycle in the CY fiber and along the $arg(\s)$ direction (the $\Th$ circle) in the base.

Even for this class of solutions, writing down an explicit form for the induced metric and flux is difficult since this requires knowledge of the function $A(\f,\bar{\f},R)$ defined in~\C{constraint}. Solving for $A$ requires finding the roots of a degree $n+2$ polynomial, which can only be solved in closed form for $n\leq2$.

To better understand these models, first consider the simplest case possible:  $n=0$. We expect the fibers to look like $\O(-1)\rightarrow\PP^0$ which is nothing more than a copy of $\CC$. Take $Y=2 \f^0\f^1$ as the gauge invariant coordinate for the fiber (the factor of 2 has been chosen for later convenience). The total space has real dimension four. In the $N\rightarrow\infty$ limit, the induced target space fields have the simple form
\bea\label{metric2}
&& ds^2 = {|dY|^2 + dR^2 + d\Th^2\over4\left(R^2+|Y|^2\right)^{1/2}}, \\
&& e^{2\varphi} = \D = {1\over 2 \left(R^2+|Y|^2\right)^{1/2}}, \\
&& H = \hlf d\Th\wedge d\Om_2, \label{H2}
\eea
where $d\Om_2$ is the volume form of the $S^2$ embedded in $(Y,\bar{Y},R)$ space. The factor of $\hlf$ appearing in $H$ is reassuring, since this guarantees that $H/2\pi$ is integrally quantized
\be
{1\over2\pi}\int H = {1\over4\pi}\int d\Th\wedge d\Om_2 = 2\pi N = Q_P.
\ee
Up to an overall factor of 2 in the metric, the fields~\C{metric2}-\C{H2}\ are precisely those for a set of $Q_P$ NS5-branes smeared over a transverse circle. It is natural to conjecture that this configuration is the endpoint of the renormalization group flow, and that this result persists even for finite values of $N$ where we expect $1/N^2$ corrections stemming from~\C{renormG}. If we pull out an overall factor of $N$ from all the coordinates, and write $R=Y_3$, then we expect the sigma-model solution to be
\bea
&& ds^2 = e^{2\varphi} \left(d\vec{Y}\cdot d\vec{Y} + d\Th^2\right), \\
&& e^{2\varphi} = 1 + {N\over 2 |\vec{Y}|}, \\
&& H = {N\over2} d\Th\wedge d\Om_2.
\eea
Notice once again that the constant FI parameters do not appear in the solutions; they have been absorbed into the fields $R$ and $\Th$. Note that the dilaton blows up at $\vec{Y}=0$ and we should not trust the string loop expansion, as usual for NS5-branes. Although the solutions we have found correspond to smeared NS5-branes, it would be interesting to see if world-sheet instantons  localize the solutions in the $\Th$ direction in a manner similar to~\cite{Tong:2002rq}.

So the simple case $n=0$ corresponds to smeared NS5-branes sitting at a point in $\R^3$, or said differently, they are wrapping a $\PP^0\subset\R^3$. A natural guess for $n>0$ is that the NS5-branes wrap the $\PP^n$ base of $\O(-n-1)\rightarrow\PP^n$, while the complex line bundle together with the $R$ direction form a transverse $\R^3$. These configurations probably cannot be realized in string theory for $n>3$, but the GLSMs are sensible nonetheless.\footnote{Even $n=3$ is subtle to interpret, since this corresponds to Euclidean NS5-branes wrapping a $\PP^3$.} A similar situation also arose in \cite{Hori:2002cd}.

Note the important difference between the $n>0$ and the $n=0$ cases (as well as the models studied in \cite{Hori:2002cd}) because the size of the $\PP^n$ varies with $R$; in particular, the $\PP^n$ has zero size at $R=0$. For $R\leq 0$, we should replace the non-linear sigma-model with a $\CC^n/\Z_n$ orbifold CFT. If $N>0$, we expect large quantum corrections in the region $R\rightarrow-\infty$, since that is where $|\s|\rightarrow 0$. In a standard GLSM, these two descriptions would appear as different ``phases" of the same theory, but now they appear within the same geometry just at different values of $R$. It would be nice to find more quantitative tests confirming that these proposed NS5-brane configurations are the endpoints of the RG flows, perhaps along the lines of~\cite{Hori:2001ax}. 


\subsection{Compact conformal models?}

There are many ways to generalize the models of section~\ref{noncompactconformal}. Take any standard GLSM and include  $\s$-dependent FI terms, while imposing $\sum_i Q^a_i=0$ for each $U(1)$.  The result should be a non-compact, non-K\"ahler $SU(n)$ structure background. Another fascinating direction is to try to build \textit{compact} models which mimic the usual hypersurface or complete intersection construction. This means introducing a superpotential and studying the zero locus. Without a superpotential, it is not possible to find conformal compact solutions because there are no positivity arguments bounding $|\s|$ in models with negatively charged fields.

We will end by describing some difficulties one encounters trying to build compact conformal models. It is useful to revisit the structure of superpotentials in $(2,2)$ models~\cite{Witten:1993yc}. For a hypersurface in a space like $\O(-n-1)\rightarrow\PP^n$, we consider a $(2,2)$ superpotential of the form
\be
\int d^2\theta \, \phi^0 W(\phi),
\ee
where $W$ is degree $n+1$ in the charge $+1$ fields, while $\phi^0$ is the distinguished field with charge $-n-1$. In $(0,2)$ superspace, this corresponding superpotential has the form
\be\label{twotwo}
S_J = -{1\over \sqrt{2}}\int\d^2x\d\th^+\, \left( \G^0 \cdot W(\phi) + \G^i \phi^0 J_i(\phi) \right)+ c.c.,
\ee
where $J_i = \partial_i W$. Here $\G^0$ is the $(2,2)$ left-moving partner of $\phi^0$. The advantage of starting with the  field content of a $(2,2)$ model is that anomaly cancelation is guaranteed to work. Note that this is a highly non-generic superpotential! Otherwise, there would be no interesting moduli space at all. 

Let us try to generalize the compact non-conformal example of section~\ref{compact}. It is useful to think about this model from the perspective of the two symmetric $D$-term constraints~\C{twodterms}. Without the $\s$-couplings,  we would have made this a conformal model  by introducing a bifundamental field $\phi^0$ with charges $(Q_0, {\tilde Q}_0)$ where:
\be
Q_0 =-  \sum_i Q_i, \qquad {\tilde Q}_0 =-  \sum_i {\tilde Q}_i, \qquad Q_i, {\tilde Q}_i>0. 
\ee
The real difficulty in finding an analogue of the complete intersection construction is writing down a superpotential which satisfies transversality. We would like to find a $W$ with charges $(-Q_0, -{\tilde Q}_0)$ under the two $U(1)$ actions such that the only solution to the $D$-term conditions and the constraints,
\be\label{Wcondition}
W= \phi^0 J_i=0,
\ee
is $\phi^0=0$ and $W=0$. This would be a possible compact conformal solution. There are additional desirable conditions  to impose on the choice of charges, described in many places like~\cite{Quigley:2011pv}. For example, we might demand a $U(1)_L$ symmetry, but let us not worry about those additional constraints at the moment.

A $W$ with this charge assignment is necessarily constructed from summing monomials of the form $\phi^n {\tilde \phi}^m$. Any interesting example will have $n>1$ or $m>1$. Taking $J_i = \partial_i W$, we see that there is a flat direction in the potential when either $\phi=0$ or ${\tilde \phi}=0$. This flat direction is usually lifted by the $D$-term constraints which, in the absence of the log interactions, force some $\phi^i$ and some ${\tilde\phi}^i$ to be non-vanishing. In our case, the $\s$-coupling permits a solution to the $D$-term constraints~\C{twodterms}\ with either all $\phi^i=0$ or all $\tilde\phi^i=0$. 

We still have a non-compact direction where $\phi^0 \neq 0$ for this attempt which does not stray very far from the structure~\C{twotwo}\ appearing in $(2,2)$ models. For more general $(0,2)$ models, there is a great deal of freedom to play with the structure of the superpotential and the choice of left-moving fermions. Increasing the number of left-moving fermions, subject to the quadratic constraint imposed by anomaly cancelation, increases the number of $J_i$ constraints. 

Even with this freedom, it is hard to lift the flat directions in the potential. Indeed, these difficulties suggest that it might not be possible to find compact conformal solutions in this class of classically gauge invariant models. If so, there should be an argument explaining the existence of flat directions from space-time physics. At this stage, we hesitate to make a stronger statement because there is a very large space of possible generalizations to explore.  Clearly, there is much to be uncovered.



\subsection*{Acknowledgements}

It is our pleasure to thank Emil Martinec and Ilarion Melnikov for helpful discussions. C.~Q. is supported in part by NSF Grant No.~PHY-0758029. S.~S. is supported in part by
NSF Grant No.~PHY-0758029 and NSF Grant No.~0529954. M.~S. is funded by NSF grant DMS 1005761.

\newpage
\appendix

\section{Superspace and Superfield Conventions}
\label{superfieldconventions}

\subsection{Chiral and Fermi superfields}

In this appendix, we summarize our notation and conventions. For a nice review of $(0,2)$ theories, see~\cite{McOrist:2010ae}. Throughout our discussion, we will use the language of $(0,2)$ superspace  with coordinates $(x^+,x^-,\th^+,\thbar^+)$. The world-sheet coordinates are defined by $x^\pm = \hlf(x^0\pm x^1)$ so  the corresponding derivatives $\del_\pm = \del_0 \pm \del_1$ satisfy $\del_\pm x^\pm =1$.
We define the measure for Grassman integration so that $\d^2\th^+ = \d\thbar^+ \d\th^+$ and  $ \int\d^2\th^+\, \th^+\thbar^+ = 1.$ The $(0,2)$ super-derivatives
\be
D_+ = \del_{\th^+} - i\thbar^+\del_+, \qquad \Dbar_+=-\del_{\thbar^+} +i\th^+\del_+,
\ee
satisfy the usual anti-commutation relations
\be
\{D_+,D_+\} = \{\Dbar_+,\Dbar_+\} = 0, \qquad \{\Dbar_+,D_+\} = 2i\del_+ .
\ee

In the absence of gauge fields, $(0,2)$ sigma models involve two sets of superfields: chiral superfields annihilated by the $\Dbar_+$ operator,
\be
\Dbar_+\F^i=0,
\ee
and Fermi superfields $\G^\a$ which satisfy,
\be
\Dbar_+\G^\a = \sqrt{2} E^\a,
\ee
where $E^\a$ is chiral: $\Dbar_+ E^\a=0$. These superfields have the following component expansions:
\bea
\F^i &=& \f^i+\sqrt{2}\th^+\j_+^i - i \th^+\thbar^+\del_+\f^i, \label{chiral}\\
\G^\a &=& \g^\a +\sqrt{2}\th^+F^\a - \sqrt{2} \thbar^+ E^\a - i\th^+\thbar^+\del_+\g^\a.
\eea

If we omit superpotential couplings, the most general  Lorentz invariant $(0,2)$ supersymmetric action involving only chiral and Fermi superfields and their complex conjugates takes the form,
\be \label{(0,2) sigma}
\L =-\hlf\int\d^2\th^+\left[{i\over2}K_i \del_- \F^i - {i\over2} K_{\ibar } \del_- {\bar \F}^{\ibar} +h_{\a\bar{\b}}\bar{\G}^{\bar \b}\G^\a + h_{\a\b}\G^\a\G^\b +h_{\bar{\a}\bar{\b}}\bar{\G}^{\bar{\a}}\bar{\G}^{\bar{\b}} \right].
\ee
The one-forms $K_i$ determine the metric; the functions $h_{\a\b}$ and $h_{\a\bar{\b}}$ determine the bundle metric.

\subsection{Gauged linear sigma models}

We now introduce gauge fields. For a general $U(1)^r$ abelian gauge theory, we require a pair $(0,2)$  gauge superfields $A^a$ and $V_-^a$ for each abelian factor, $a=1,\ldots,r$. Let us restrict to $r=1$ for now. Under a super-gauge transformation, the vector superfields transform as follows,
\bea
\dd A &=& {i}(\bar{\La} - \La)/2, \\
\dd V_- &=& - \del_-(\La + \bar{\La})/2,
\eea
where the gauge parameter $\La$ is a chiral superfield: $\Dbar_+ \La=0$.  In Wess-Zumino gauge, the gauge superfields take the form
\bea
A &=& \th^+\thbar^+ A_+, \\
V_- &=& A_- - 2i\th^+\labar_- -2i\thbar^+\la_- + 2\th^+\thbar^+ D,
\eea
where $A_\pm = A_0 \pm A_1$ are the components of the gauge field. We will denote the gauge covariant derivatives by
\be
\cD_\pm = \del_\pm + i Q A_\pm
\ee
when acting on a field of charge $Q$. This allows us to replace our usual superderivatives $D_+,\Dbar_+$ with gauge covariant ones
\be
\mathfrak{D}_+ = \del_{\th^+} - i\thbar^+\cD_+ \qquad \bar{\mathfrak{D}}_+=-\del_{\thbar^+} +i\th^+\cD_+
\ee
which now satisfy the modified algebra
\be
\{\mathfrak{D}_+,\mathfrak{D}_+\} = \{\bar{\mathfrak{D}}_+,\bar{\mathfrak{D}}_+\} = 0 \qquad \{\bar{\mathfrak{D}}_+,\mathfrak{D}_+\} = 2i\cD_+ .
\ee
We must also introduce the supersymmetric gauge covariant derivative,
\be
\nabla_- = \del_- + i Q V_-,
\ee
which contains $\cD_-$ as its lowest component. The gauge invariant Fermi multiplet containing the field strength is defined as follows,
\be
\Upsilon =[\bar{\mathfrak{D}}_+,\nabla_-] =  \Dbar_+(\del_- A + i V_-) = -2\big(\la_- - i\th^+(D-iF_{01}) - i\th^+\thbar^+\del_+\la_-\big).
\ee
Kinetic terms for the gauge field are given by
\be \label{LU}
\L = {1\over8e^2}\int\d^2\th^+\, \bar{\Upsilon}\Upsilon = {1\over e^2}\left(\hlf F_{01}^2 + i\labar_-\del_+\la_- + \hlf D^2\right).
\ee
Since we are considering abelian gauge groups, we can also introduce an FI term with complex coefficient $t=ir + {\th\over2\pi}$:
\be \label{LFI}
{t\over4}\int\d\th^+ \Upsilon\Big|_{\thbar^+=0} + c.c. = -rD + {\th\over2\pi}F_{01}.
\ee

In order to charge our chiral fields under the gauge action, we should ensure that they satisfy the covariant chiral constraint $\mathfrak{\bar{D}}_+\Phi = 0$. Since $\mathfrak{\bar{D}}_+ = e^{QA}\Dbar_+e^{-QA}$ it follows that $e^{QA}\Phi_0$ is a chiral field of charge $Q$, where $\Phi_0$ is the neutral chiral field appearing in \C{chiral}. In components,
\be
\F = \f + \sqrt{2} \j -i\th^+\thbar^+\cD_+\f
\ee
The standard kinetic terms for charged chirals in $(0,2)$ gauged linear sigma models (GLSMs) are
\bea \label{LPhi}
\L &=& {-i\over2}\int\d^2\th^+\ \bar{\F}^i \nabla_- \F^i, \\
&=& \left(-\big|\cD_\mu \f^i\big|^2 + \bar{\psi}_+i\cD_-\psi_+^i - \sqrt{2}iQ_i \bar{\f}^i\la_-\j^i_+ + \sqrt{2}iQ_i\f^i\bar{\j}_+^i\labar_- + D Q_i \big|\f^i\big|^2\right). \non
\eea
Fermi superfields are treated similarly. We promote them to charged fields by defining $\G = e^{QA}\G_{0}$ so that in components
\be
\G = \g + \sqrt{2}\th^+F + \sqrt{2}\thbar^+E -i\th^+\thbar^+\cD_+\g.
\ee
If we make the standard assumption that $E$ is a holomorphic function of the $\F^i$ then the  kinetic terms for the Fermi fields are:
\bea \label{LLa}
\L &=& -\hlf\int\d^2\th^+\,  \bar{\G}^\a \G^\a, \\
&=& \left(i\bar{\g}^\a\cD_+\g^\a + \big|F^\a\big|^2 - \big|E^\a\big|^2 - \bar{\g}^\a\del_i E^\a \j_+^i - \bar{\j}_+^i \del_\ibar \bar{E}^\a \g^\a\right). \non
\eea

\subsection{Superpotential couplings}

We can introduce superpotential couplings,
\be\label{super}
S_J = -{1\over \sqrt{2}}\int\d^2x\d\th^+\, \G \cdot J(\F) + c.c.,
\ee
supersymmetric if $E\cdot J=0$, which give a total bosonic potential
\be
V = |E|^2 + |J|^2.
\ee
The action consisting of the terms \C{LU},~\C{LFI},~\C{LPhi},~\C{LLa}\ and~\C{super}\ comprises the standard $(0,2)$ GLSM.

\section{Proof that $\left[{\cal R}^{(-)}\right]=0 \, \Leftrightarrow \,\sum_i Q^a_i=0$}
\label{proof}

In section~\C{obs}\ we noted that the only obstruction to the triviality of ${\cal R}^{(-)}$ in cohomology is that $\del\delbar\log\det G_{i\jbar}$ be a trivial class. This follows if the first Chern class of the fiber is trivial. We would like to show that this condition corresponds to $\sum_i Q^a_i=0$ for each $U(1)$ factor of the gauge group in the GLSM.



Recall that in projective coordinates, the fiber metric takes the form
\be
G_{i\jbar} = e^{2Q_i^cA^c}\left(\dd_{i\jbar}-2\bar{\f}_i\f_\jbar Q^a_i \D^{ab} Q_j^b e^{2Q_j^cA^c}\right),
\ee
with $A^a$ defined implicitly via~\C{constraint}, and $\D^{ab}$ is given in~\C{delta}. In terms of these projective coordinates, $\det G_{i\jbar}$ vanishes so we must work in a local patch using gauge invariant coordinates. First we must generalize the local coordinates~\C{coords}\ suitably for higher rank gauge groups.\footnote{We found a similar discussion in \cite{Adams:2006kb} useful for these definitions.} Each patch must now be labeled by a multi-index $\cA\subset\{1,2,\ldots,n\}$ with $|\cA|=r$. For $\a\in\cA$ we require that the $r\times r$ matrix $Q^a_\a$ be invertible. We then define the patch $U_{(\cA)}=\left\{\f^i\in\CC^n\big|\f^\a\neq0,\forall \a\in\cA\right\}$. Within that patch, we can define the coordinates:
\be
Z^i_{(\cA)} = \f^i \prod_{\a\in\cA} \left(\f^\a\right)^{-(Q^{-1})^\a_a Q^a_i}.
\ee
The transformation properties on intersections $U_{(\cA)}\cap U_{(\B)}$ for the coordinates $Z_{(\cA)}$ and the sections $A_{(\cA)}$ are easy enough to work out,
\bea
Z^i_{(\cA)} &=& Z^i_{(\B)}\prod_{\a\in\cA}\left(Z^\a_{(\B)}\right)^{-(Q^{-1})^\b_a Q^a_i}, \\
A^a_{(\cA)} &=& A^a_{(\B)} + \sum_{\a\in\cA}(Q^{-1})^\a_a \log\left|Z^\b_{(\cA)}\right|.
\eea
A straightforward but somewhat involved calculation reveals that,
\be
\log\det G_{i\jbar} = 2\left(\sum_{i}Q^a_i\right)A_{(\cA)}^a + 2\log\det Q^a_\a + \log\det\left(2\D^{ab} \right).
\ee
The functions $\D^{ab}$ are globally defined, but as we see from the transformation properties above, $A^a$ is a non-trivial section of some line bundle $\L^a$.  The only way to ensure that $\del\delbar\log\det G_{i\jbar}$ is trivial in cohomology is to impose $\sum_i Q^a_i=0$ for each $U(1)$ factor.

\section{The trouble with $S^3\times S^1$}
\label{S3xS1}

The $SU(2)\times U(1)$ WZW models are a well studied family of conformal field theories associated with a compact non-K\"ahler manifold. See, for example,~\cite{Spindel:1988nh, Spindel:1988sr, Rocek:1991vk}.  The target space is $S^3\times S^1$, which we can view as $E\rightarrow \PP^1$, where $E$ is a torus constructed from the Hopf fiber of $S^3\cong S^1\rightarrow S^2$ together with the free circle. See, for example,~\cite{Witten:2005px}. The family of conformal field theories is labeled by the amount of integer $H$-flux threading the $S^3$.

Let $(\f,\th)$ be coordinates on the Hopf fiber and the free circle, respectively, and let $z=\f+i\th$ be a complex combination parameterizing $E$.  Note that $z$ is not a complex coordinate on $S^3\times S^1$ because the complex structure operator maps $d\th$ to $d\phi+A$, where $A$ is the potential for the K\"ahler form on $\PP^1$: $dA=J_{FS}$. In these coordinates, the fundamental form for the space is given by
\be
J = J_{FS} +i(dz+A)\wedge(d\bar{z}+A).
\ee
The $H$-flux threading the $S^3$ takes the form
\be
H = \hlf\left( dz+A\right)\wedge J_{FS} + c.c.,
\ee
where we have assumed one unit of flux.
A non-linear sigma model with this target space is a perfectly good CFT; however, this is not an admissible string background because there is no well-defined dilaton. In particular, the Lee form
\be
\xi + {\bar \xi} = {i\over 2}\left( d\bar{z} + A\right) - {i\over2}\left(dz+A\right) = d\th
\ee
is not exact. Only when we replace $S^1 = {\mathbb R}/{\mathbb Z}$ by its cover ${\mathbb R}$ does $\th$ becomes a globally defined function. With this replacement, it make sense to identify $\th\sim2\varphi$.

This should be contrasted with all the models constructed in this paper. Despite the non-trivial circle factors in our solutions,
the Lee form found in the GLSM never has components along those directions. This is evidence for the assertion that the IR fixed points of these theories can be used to construct heterotic string backgrounds.

\newpage


\ifx\undefined\bysame
\newcommand{\bysame}{\leavevmode\hbox to3em{\hrulefill}\,}
\fi

\end{document}